\documentclass[%
 preprint,
 superscriptaddress,
 amsmath,amssymb,
 aps,
 prb,
]{revtex4-2}

\usepackage{graphicx}
\usepackage{ amssymb }
\usepackage{placeins}
\usepackage{dcolumn}
\usepackage{bm}
\usepackage{mathrsfs}
\usepackage{hyperref}
\usepackage{color}
\hypersetup{colorlinks=true,linkcolor = blue,
            urlcolor  = blue,
            citecolor = blue,
            anchorcolor = blue, linktocpage}

\usepackage[caption=false]{subfig}
\DeclareUnicodeCharacter{2212}{-}

\begin{document}

\title{Symmetry-dependent antiferromagnetic proximity effects on valley splitting}

\author{Chengyang Xu}
\affiliation{Key Laboratory of Artificial Structures and Quantum Control (Ministry of Education), School of Physics and Astronomy, Shanghai Jiao Tong University, Shanghai 200240, China}

\author{Lingxian Kong}
\affiliation{International Center for Quantum Materials, School of Physics, Peking University, Beijing 100871, China}

\author{Weidong Luo}%
\email{wdluo@sjtu.edu.cn}
\affiliation{%
Key Laboratory of Artificial Structures and Quantum Control (Ministry of Education), School of Physics and Astronomy, Shanghai Jiao Tong University, Shanghai 200240, China}%
\affiliation{Institute of Natural Sciences, Shanghai Jiao Tong University, Shanghai 200240, China}

\date{\today}

\begin{abstract}
Various physical phenomena have been discovered by tuning degrees of freedom, among which there is the degree of freedom (DOF) −- ``valley". The typical valley materials are characterized by two degenerate valley states protected by time-reversal symmetry ($\rm{\mathcal{T}}$S). These states indexed by valley DOF have been measured and manipulated for emergent valley-contrasting physics with the broken valley degeneracy. To achieve the valley splitting resulted from $\rm{\mathcal{T}}$S breaking, previous studies mainly focused on magnetic proximity effect provided by ferromagnetic (FM) layer. In contrast, the anti-ferromagnetic (AFM) proximity effect on the valley degeneracy has never been investigated systematically. In this work, we construct the composite systems consisting of a transition-metal dichalcogenide (TMD) monolayer and a proximity layer with specific intra-plane AFM configurations. We extend the three-band model to describe the valley states of such systems. It is shown that either ``time-reversal + fractional translation" or ``mirror" symmetry can protect the valley degeneracy. Additionally, first-principles calculations based on density functional theory (DFT) have been performed to verify the results obtained from the extended tight-binding (TB) model. The corresponding mechanism of the valley splitting/degeneracy is revealed through the non-degenerate perturbation. Meanwhile, an extra condition is proposed to keep the well-defined valley states disentangled with each other through two negative examples based on degenerate perturbation. Further DFT studies on the effects of the $U_{\mathrm{eff}}$ and interlayer distance are performed. Manipulating the magnetization of Mo is shown to be feasible and effective for controlling the valley splitting with the direct overlap tuned by $U_{\mathrm{eff}}$ and the interlayer distance. The TB method introduced in the present work can properly describe the low-energy physics of valley materials that couple to the proximity with complex magnetic configurations. The results considerably expand the range of qualified proximity layers for valley splitting, enabling more flexible manipulation of valley degree.
\end{abstract}

\maketitle

\section{\label{Introduction}Introduction}
2D systems provide ideal platforms for exploring physical phenomena by modifying different DOF. In some 2D materials, especially the H-phase monolayer TMDs, there is the DOF $--$ ``valley" for electrons in the low-energy region \cite{xiao2007valley,yao2008valley,cao2012valley,xiao2012coupled}. As the first proposed monolayer in TMD family, strong valley-selective photoluminescence in directly gapped MoS$_2$ was observed with optical pumping\cite{exciton1,Zeng2012}. Subsequent investigations into valley-contrasting physical quantities have been reported in many other TMDs (e.g. MoSe$_2$, WS$_2$, and WSe$_2$). Electrons with different valley indices can be controllably excited through the helicity of light \cite{lighthelicity1,lighthelicity2}. With in-plane electric field, valley degree can be detected through the valley Hall effects where carriers at different valleys transport along opposite boundaries \cite{xiao2007valley,valleyhall}. Analogous to spintronics, electronic valley as an information carrier may open up a new paradigm for data processing and storage based on the understanding and precise control of the valley DOF \cite{datastorage1,datastorage2}.\par

\begin{figure}
\includegraphics[width=17cm]{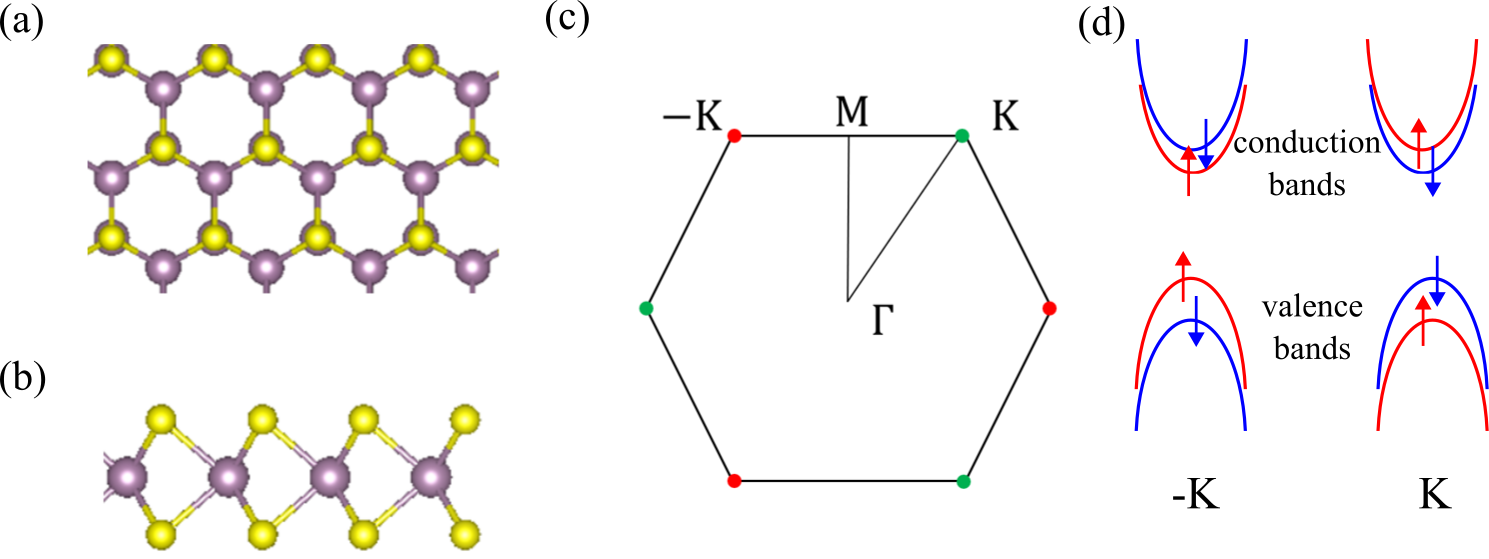}%
\caption{(a), (b) Top and side view of the valley material monolayer TMD MX$_2$ (M=Mo, W; X=S, Se). The purple/yellow spheres denote the M/X atoms.
(c) Brillouin zone and high-symmetry points of monolayer TMD, with red and green points denoting the two valleys. (d) Schematic conduction and valence bands of MX$_2$ at two valleys. Spin-up/spin-down are denoted by red/blue arrows. The spin orientations of the two eigenstates with the same energy are opposite at K and $-$K due to the $\rm{\mathcal{T}}$S.}
\label{1}
\end{figure}

As an intrinsic nature of materials, the valley degree is strongly related to the symmetry of the system. One of the most promising 2D materials for manipulating valley degree is the H-phase monolayer TMDs with the direct gap locating at the corners of the hexagonal Brillouin zone (BZ). This triangular type lattice consists of pristine cells with one transition-metal (M) atom centered in the prism formed by the nearest chalcogen (X) atoms as shown in Fig.~\ref{1}(a)--(b). The loss of inversion symmetry and strong spin-orbit coupling lead to the spin-valley coupled states at the band edges K and $-$K. The corresponding spin orientation at two valleys are opposite due to the time-reversal symmetry as schematically shown in Fig.~\ref{1}(d). This spin-valley locking relationship is the key to the observable effects such as valley Hall effect \cite{valleyhall}, valley-contrasting circular dichroism \cite{xiao2012coupled} and valley-selective excitation of excitonic states \cite{exciton1,exciton2,exciton3}.\par

Recently, lifting the valley degeneracy protected by $\rm{\mathcal{T}}$S has been implemented for precise control of the valley effects in such monolayer TMDs. An external magnetic field was utilized to induce a small valley splitting of $0.1$-$0.2$ meV/T in previous works \cite{magnetic1,magnetic2,magnetic3}. In comparison, magnetic proximity effect could generate giant valley splitting, increasing by an order of magnitude to several meV/T, in semiconductor/ferromagnet hybrids \cite{seyler2018valley,qi2015giant,prl.124.197401}. A sizable valley splitting also occurs under the AFM proximity effect with A-type configuration \cite{yang2018induced}. And the exchange field was found to be dominated by the inter-facial magnetic layer, where the magnetic moments are FM coupling. Hence, the proximity effect is still limited to the case of the FM type. \par

So far, it is still unclear whether intra-plane AFM proximity would result in the valley splitting of TMD monolayers. The systems seem to be valley degenerate under the intra-plane AFM proximity effect since the exchange field over the whole magnetic unit cell is zero. However, the orbital-dependent exchange paths and distance-dependent magnetic exchange yield a nonzero effective Zeeman splitting for each atom. Thus, it is hard to determine whether the valley degeneracy breaks in terms of the overall effective exchange field. Yet, the symmetry of the proximity layer offers an effective method to analyze the problem. If the valley degeneracy remains under the magnetic proximity effect, there must exist at least one symmetry that relates the two states featured by (K,$\sigma$) and ($-$K,$\overline{\sigma}$).\par 

In the present paper, we start from the symmetries that can reverse the spin and momentum simultaneously. Three special AFM configurations are constructed in Sec.~\ref{sec2A}. In Sec.~\ref{sec2B}, MoTe$_2$ monolayer has been adopted as the TMD layer. A three-band TB model is extended with an additional term to study the proximity effects of the 3 specific AFM configurations. In Sec.~\ref{sec2C}, first-principles DFT calculations are carried out to validate the TB results. In Sec.~\ref{3A}, the vertical ``mirror" ($\rm{\sigma_v}$) symmetry in the type \uppercase\expandafter{\romannumeral2} and the ``time-reversal + fractional translation" ($\rm{\mathcal{T}t_R}$) symmetry in type \uppercase\expandafter{\romannumeral1} are shown to protect the valley degeneracy through the symmetry analysis. And the valley splitting appears as predicted in the type \uppercase\expandafter{\romannumeral3}, exhibiting the dependence on orbital-resolved exchange field. In Sec.\ref{3B}, the non-degenerate perturbation theory is adopted to explain the mechanism of the valley splitting/degeneracy in the 3 types of AFM configurations. In order to preserve the well-defined valley states, another 2 special configurations, type \uppercase\expandafter{\romannumeral4} and type \uppercase\expandafter{\romannumeral5}, are constructed as the negative examples. These two configurations are featured by the reciprocal lattice vectors through which K and $-$K fold to the same point. Based on the degenerate perturbation theory, the constraints of the AFM configurations are obtained. In Sec.~\ref{3C}, the dependence of the valley splitting on the interlayer distance and $U_{\mathrm{eff}}$ parameter has been studied through the DFT calculations. A larger valley splitting is induced with a larger overlap between orbitals of the M atoms and the magnetic ones. The magnetization of the M atom is found to be the key factor tuned through the interlayer distance and $U_{\mathrm{eff}}$.\par

\section{Methods}

\subsection{\label{sec2A}Symmetry analysis for constructing AFM configurations}
The degenerate valley states refer to the two states, indexed by different valleys, holding the same eigen-energy non-accidentally. In MX$_2$ (Fig.~\ref{1}), the valleys locate at K and $-$K where bands reach the valence band maximum and conduction band minimum. The $\rm{\mathcal{T}}$S resulting from the nonmagnetism protects the Kramers' degeneracy of the two states specified by opposite momentum and spin orientation. When the magnetic proximity effect is introduced, the $\rm{\mathcal{T}}$S is broken. If the normal degeneracy of valley states remains, at least one common symmetry other than $\rm{\mathcal{T}}$S of the TMD monolayer and proximity layer reverses the momentum and spin indexes of the valley states.\par

\begin{figure}
\includegraphics[width=17cm]{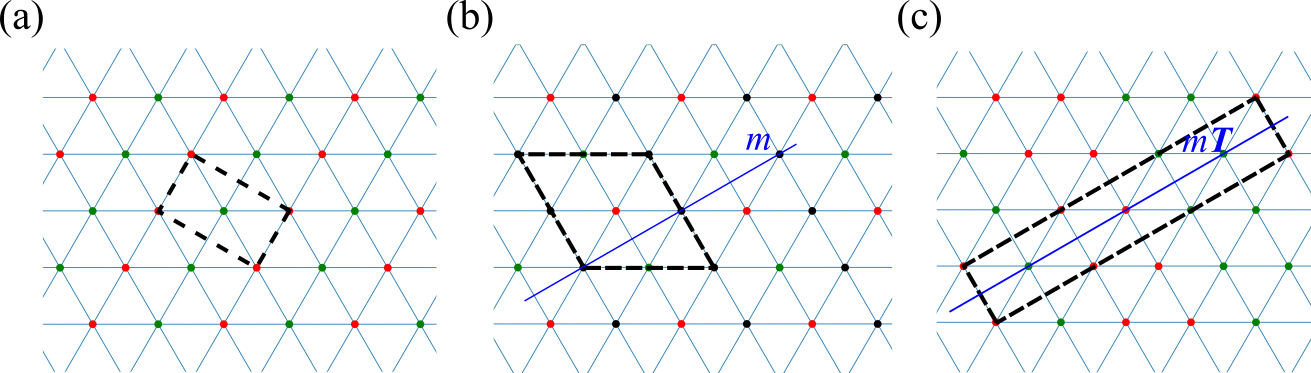}%
\caption{(a)--(c) Type \uppercase\expandafter{\romannumeral1}--\uppercase\expandafter{\romannumeral3} intra-plane AFM configurations. Red/green points denote the nearest neighboring exchange field provided by the spin magnetic moment $\uparrow$/$\downarrow$. The spin-quantization axis is normal to the layer. Black points denote non-proximity effect at the corresponding sites. Black dashed lines denote the unit cell after intra-plane AFM proximity is introduced. (b) The type \uppercase\expandafter{\romannumeral2} is symmetric about the vertical mirror plane labeled by the blue line. (c) The type \uppercase\expandafter{\romannumeral3} is unchanged under the vertical mirror symmetry combined with $\rm{\mathcal{T}}S$, which is not the group element of the TMD layer.}
\label{f2}
\end{figure}

Based on the symmetry analysis above, two hypothetical AFM configurations are constructed in Fig.~\ref{f2}(a)--(b). The configuration in Fig.~\ref{f2}(a) keeps the $\rm{\mathcal{T}t_R}$ symmetry. After the intra-plane AFM proximity taken into account, the enlarged unit cell including two magnetic atoms in real space corresponds to a downsized irreducible BZ in Fig.~\ref{f7}(a). Noted that the fractional translation of the periodic AFM configuration is a linear combination of integer multiple of the lattice vectors of the MX$_2$. The $\rm{\mathcal{T}t_R}$ converts the red points (sites with net spin-up) to the green ones (sites with net spin-down) through the $\rm{t_R}$ and then swaps them again with $\rm{\mathcal{T}}$. The other configuration (Fig.~\ref{f2}(b)) holds the $\rm{\sigma_v}$ symmetry enabling the flipping of the spin $z$ component. The $2\times2$ construction of the unit cell generates a smaller hexagonal BZ without any rotation as shown in Fig.~\ref{f7}(b). Both AFM configurations in Fig.~\ref{f2}(a)--(b) are expected to keep the valley degeneracy. It’s worth mentioning that both magnetic configurations of type \uppercase\expandafter{\romannumeral1} and type \uppercase\expandafter{\romannumeral2} hold the ``inversion" symmetry. However, when they couple to the TMD monolayer, the inversion symmetry will never be the group element of the whole system. On the other hand, the configuration with neither $\rm{\mathcal{T}t_R}$ or $\rm{\sigma_v}$ symmetry (Fig.~\ref{f2}(c)) is considered as well. It aims to illustrate that the absence of the non-trivial symmetries of the coupled system is responsible for valley splitting. For abbreviation, the valley splitting in the following parts refers to the splitting of the two degenerate valley states defined in the TMD monolayer. It is quantified by energy difference of the top valence band at K and $-$K, which is given by $\Delta_{tvb} = E_{tvb}(\mathrm{K},\downarrow) − E_{tvb}(-\mathrm{K},\uparrow)$. The splitting of the lowest conduction band at two valleys is not included since the results and analyses follow the same pattern.\par

\subsection{\label{sec2B}Tight-binding model}
\begin{figure}
\includegraphics[width=17cm]{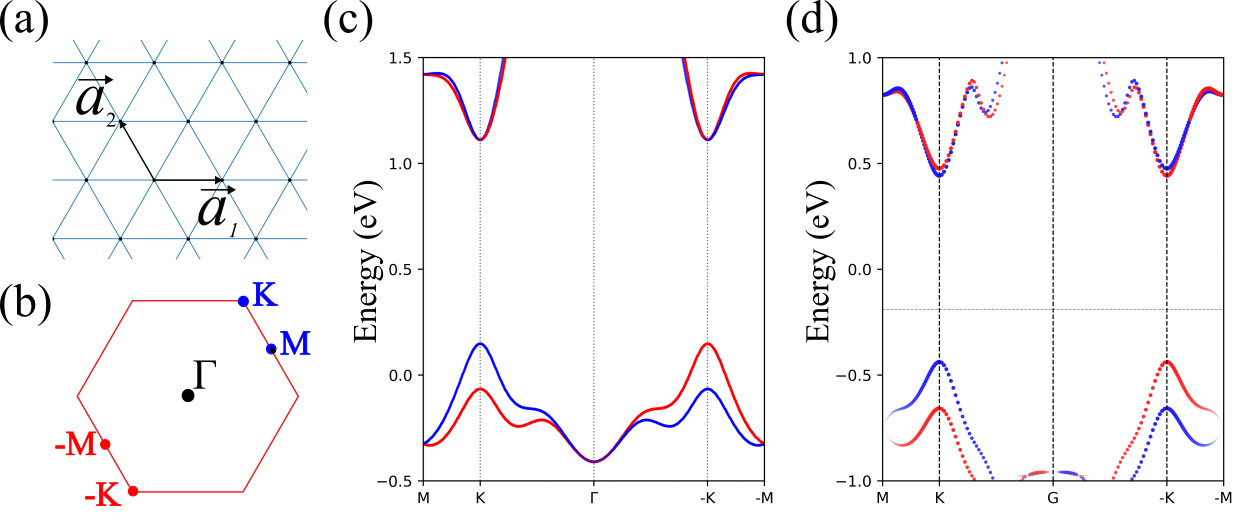}%
\caption{(a) The triangular lattice formed by M atoms of monolayer MX$_2$.
(b) Band structure of MoTe$_2$ monolayer obtained from the three-band TB model. Blue(Red) color denotes the spin-down(spin-up) channel.
(c) Projected band structure of the freestanding MoTe$_2$ monolayer from DFT calculation. The shade of the blue(red) circle denotes the weight of the projection on spin-down(spin-up) \{$d_{z^{2}}$, $d_{xy}$, $d_{x^{2}-y^{2}}$\} orbitals.}
\label{f3}
\end{figure}

Generally speaking, valley degeneracy or splitting is confirmed by eigenvalues of the bands at two valleys. Based on a three-band TB model for monolayer TMDs \cite{liu2013three}, the conduction and valence bands at $\pm$K valleys are well-described in the minimal basis composed of Mo \{$d_{z^{2}}$, $d_{xy}$, $d_{x^{2}-y^{2}}$\}. The three-band Hamiltonian is expressed as \cite{liu2013three}
\begin{equation}
H_0 = \sum_{i,\alpha,\beta,\sigma}\mu_{\alpha \beta}c_{i\alpha\sigma}^{\dagger}c_{i\beta\sigma} + \sum_{i,j,\alpha,\beta,\sigma}h_{ij}^{\alpha\beta}c_{i\alpha\sigma}^{\dagger}c_{j\beta\sigma}
\label{eq1}
\end{equation}
with $\alpha (\beta)$, $\sigma$ denoting orbital and spin indices. And $i,j$ represent neighboring lattice sites. It should be noted that the on-site energy matrix $\mu_{\alpha\beta}$ is not diagonal because of the spin-orbit coupling. With $D_{3 h}$ point group taken into account, the number of independent parameters reduces to 9. In this work, the TB model is parameterized from the MoTe$_2$ monolayer \cite{liu2013three}. The band structure from this model is consistent with DFT calculation as shown in Fig.~\ref{f3}(c)--(d).\par
To involve the intra-plane AFM proximity effect, the effective exchange field provided by the magnetic proximity is treated as a staggered on-site energy modifier term \cite{zhu2017two}, which is given by
\begin{equation}
    H' =\sigma_z \sum_{i,\alpha,\sigma}O_{i\alpha}c^{\dagger}_{i\alpha\sigma}c_{i\alpha\sigma}.
\end{equation}
This diagonal on-site correction $O_{i\alpha}$ represents the energy shift of orbital $\alpha$ at lattice site $i$. The unfolded bands are required to remain the main features under $\vec{O}_i \neq \vec{0}$ such that the proximity effect could be viewed as a perturbation to the three-band model. The orbital-resolved exchange fields provided by spin-up and spin-down magnetic moment are opposite with $-\vec{O}_{G} = \vec{O}_{R} = \vec{O}$ as shown in Fig.~\ref{f2}. The zero matrix $\vec{O}_{B} = (0,0,0)$ eV acts on the non-proximity sites denoted by black circles in Fig.~\ref{f2}.\par

The band structures of the coupled systems are accessible with the extended three-band Hamiltonian $H=H_0+H'$. The enlarged unit cell results in a smaller BZ and more bands from the folding process, which increases the difficulty of locating the original valley states. For clarity, effective bands in the primitive BZ are obtained via unfolding technique \cite{ku2010unfolding,popescu2012extracting,farjam2015projection,huang2014general} with the unfolding projection weight $W_{\mathbf{K}\rightarrow\mathbf{k}}=\left\langle\mathbf{K} \left|\mathcal{P}_{\mathbf{k}}\right| \mathbf{K} \right\rangle$ \cite{farjam2015projection}.\par

\subsection{\label{sec2C}First-principles calculations}
We perform the first-principles calculations using Vienna Ab initio Simulation Package (VASP) \cite{vasp1,vasp2} with projector augmented-wave (PAW) method \cite{PAW} and Perdew-Burke-Ernzerhof (PBE) functional \cite{PBE1,PBE2} of the generalized-gradient approximation. In addition to spin-orbit coupling, correction of Hubbard $U$ with the rotationally invariant approach for interacting localized Mn 3$d$ orbitals \cite{+Umethod} is included in the calculations. The magnetic moments are forced to align perpendicular to the layer. Based on the DFT results, the post-processed band unfolding calculation is carried out with the plane-wave basis method \cite{popescu2012extracting}.\par

With 1H-type MoTe$_2$ as the TMD layer, the MnO (111) monolayer terminated with Mn atoms right below Mo atoms is taken as the proximity layer for two reasons. (\romannumeral1) The additional bands are disentangled from the original valley states within low-energy region. (\romannumeral2) The effective magnetic field from the half-filled 3$d$ orbitals of Mn$^{2+}$ acts on all 4$d$ orbitals of Mo. A sizable valley splitting in the case of the FM type is expected. Mn atoms are replaced with Mg atoms at non-magnetic sites in the type \uppercase\expandafter{\romannumeral2} and type \uppercase\expandafter{\romannumeral4} to keep the valence state of Mn$^{2+}$ ions unchanged.\par

Dynamic stability and lattice mismatch are neglected for the MoTe$_2$/MnO hybrid since it's not the main point of this work. The structures of AFM types are transformed from the FM MoTe$_2$/MnO primitive cell that has been relaxed in advance with the lattice constant 3.56 \r{A} \cite{qi2015giant}. In order to show that the effect of magnetic configuration on interlayer distance is negligible, the FM type, AFM type \uppercase\expandafter{\romannumeral1} and AFM type \uppercase\expandafter{\romannumeral3} are relaxed with only MoTe$_2$ layer fixed. The van der Waals interaction is included through the optB88-vdW method \cite{vdw}. Further calculations are performed to study the effect of $U_{\mathrm{eff}}$ and Mn-Mo distance $d$ on the valley splitting. The $U_{\mathrm{eff}}$ ranges from 4 to 7 eV with the ``$2+$" valence state of Mn taken into account \cite{Mn+U1,Mn+U2}. The $d$ is tuned from 3.6 to 4.8 \r{A}. For one specific $U_{\mathrm{eff}}$ and $d$, the FM-type structures are partially relaxed through fixing the $d$ and the pre-optimized MoTe$_2$ monolayer. The structures of the AFM types with the same $U_{\mathrm{eff}}$ and $d$ are then constructed by enlarging the optimized unit cell of the FM type. Noted that the Mn-Mo distances are the same in each AFM type. It implies the exactly total zero magnetization of the Mo 4$d$ orbitals over the supercell. The systems listed in Table.~\ref{tab3} and ~\ref{tab4} keep semiconducting with Mn exhibiting about 5 $\mu_B$ spin magnetic moment. A 20 \r{A} vacuum space along $z$ direction is applied to avoid periodic slab interaction in all DFT calculations. It's worth mentioning that the Mo $\{d_{z^2},\ d_{xy},\ d_{x^2-y^2}\}$ remain to be the dominant orbital components of the top valence band and the lowest conduction band at two valleys in all cases.\par
\begin{figure}
\includegraphics[width=17cm]{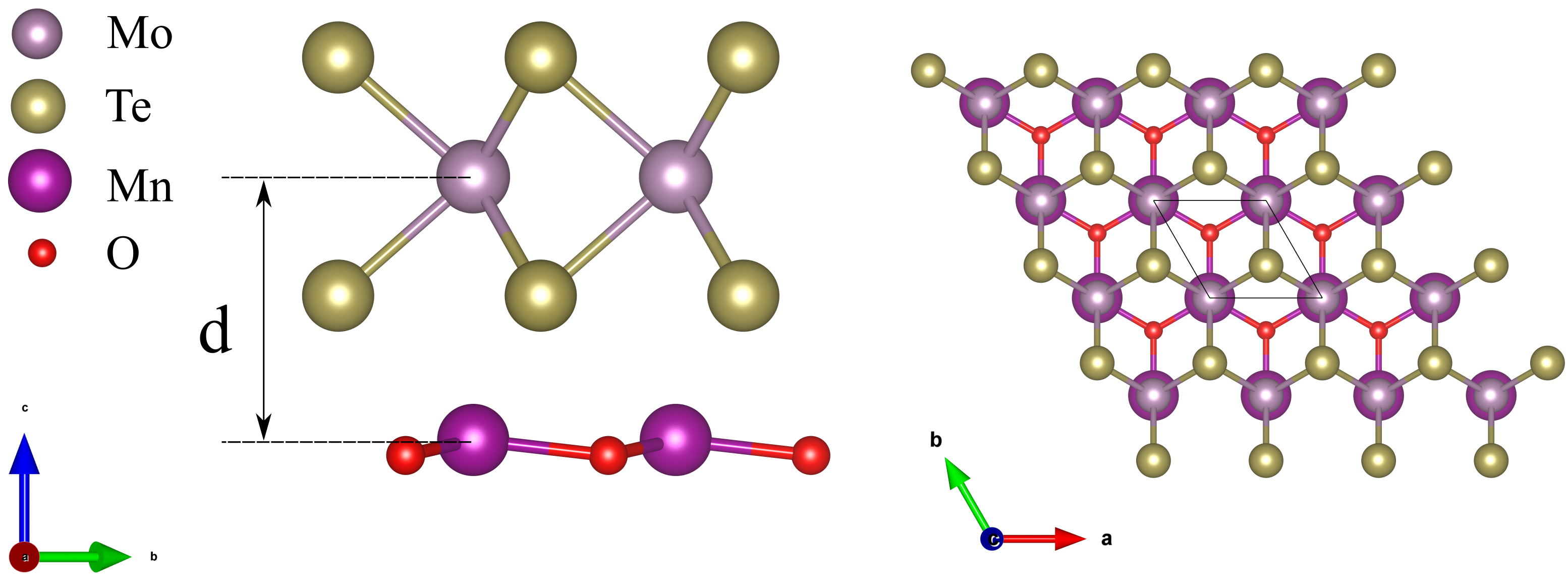}%
\caption{The side and top views of the MoTe$_2$/MnO (111) hybrid.}
\label{struct}
\end{figure}

\section{\label{discuss}Results and Discussion}
\subsection{\label{3A} The valley degeneracy and the orbital-dependent valley splitting}
\begin{figure}
\includegraphics[width=17cm]{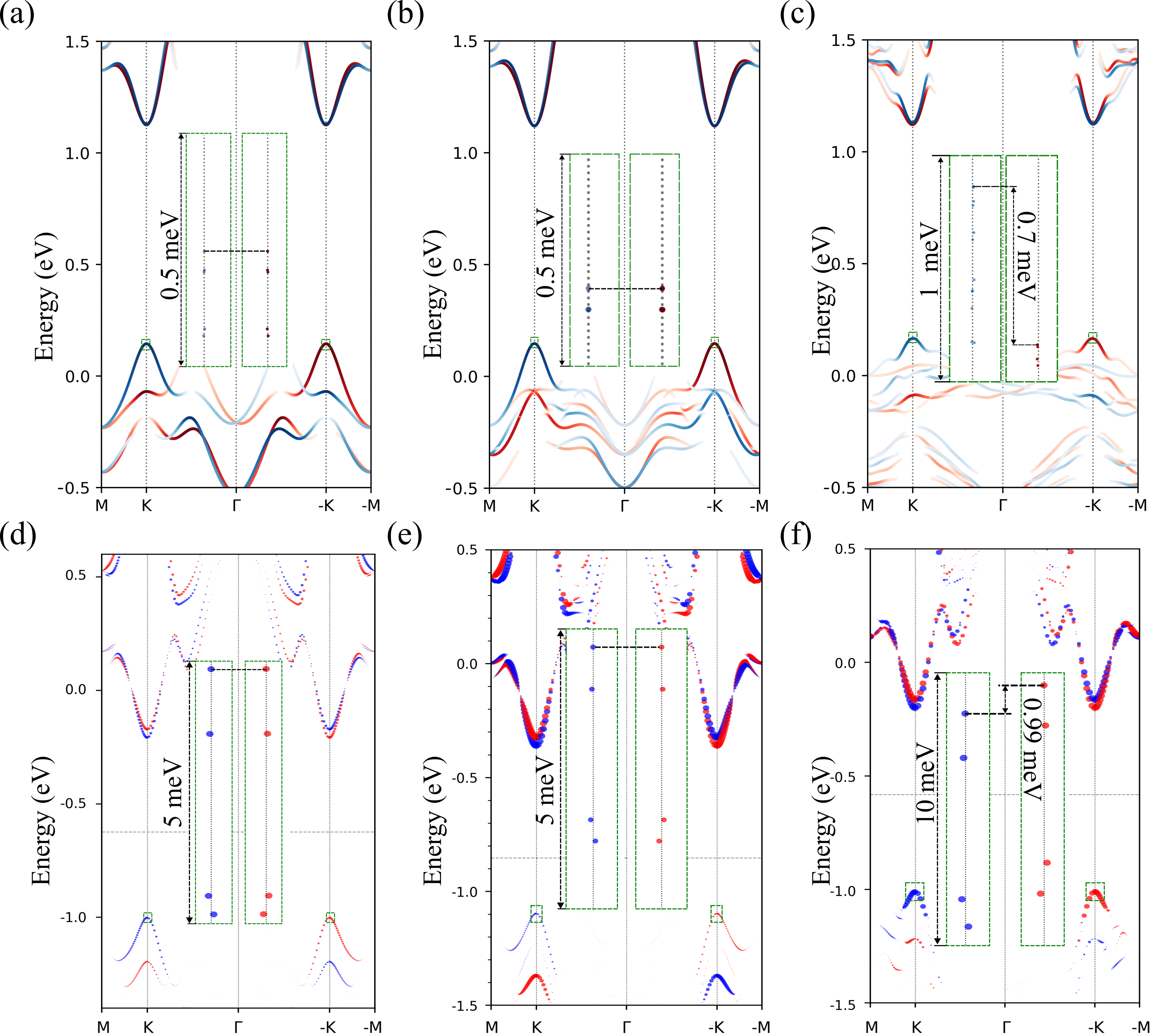}%
\caption{Unfolded bands of the type \uppercase\expandafter{\romannumeral1}, \uppercase\expandafter{\romannumeral2}, \uppercase\expandafter{\romannumeral3} hybrids. The green dashed frames represent amplified regions. (a), (c), (e) Unfolded bands from the TB models with $\vec{O} = (0.3, 0.1, 0.2)$ eV. The Red/blue circles denote unfolded bands with spin $\uparrow$/$\downarrow$. For clarity, only the points with projection weight larger than 3\% are plotted. (b), (d), (f) The unfolded bands from the DFT calculations with $U_{\mathrm{eff}}$ = 4 eV and $d$ = 3.6 \r{A}. The shade and size of the red/blue circles denote the unfolding weight multiplied by the weight of projection on Mo \{$d_{z^{2}}$, $d_{xy}$, $d_{x^{2}-y^{2}}$\} with positive/negative  component.}
\label{f5}
\end{figure}

In the type \uppercase\expandafter{\romannumeral1}, $\rm{\mathcal{T}t_R}$ symmetry reverses the spin $z$ component and the sign of momentum, transforming the quantum state $\psi_{k,\uparrow}$ to $\psi_{-k,\downarrow}$ with a translation induced phase shift. As expected, the Kramers' degeneracy remains throughout the whole BZ in the unfolded bands in Fig.~\ref{f5}(a) and Fig.~\ref{f5}(d). The valley splitting is prohibited in the type \uppercase\expandafter{\romannumeral1} even the $\rm{\mathcal{T}}$S is absent.\par
The two valleys stay degenerate in the unfolded bands in Fig.~\ref{f5}(b) and (e) although $\rm{\mathcal{T}{t}_R}$ does not belong to the group of the type \uppercase\expandafter{\romannumeral2}. The remaining degeneracy is protected by $\rm{\sigma}_v$ symmetry, which swaps green and red sites in Fig.~\ref{f3}(b). And the colors denoting spin $z$ components commute as well since the spin $z$, as a pseudo-vector, is parallel to the vertical mirror plane. Thus, the system will change back to its initial configuration in the real space. When it comes to the reciprocal space, the K of the primitive BZ transforms to $-$K after the $\rm{\sigma_v}$ operation in the BZ of MoTe$_2$ monolayer. Yet, whether the states at original valleys are degenerate depends on the states labeled by folded points from K and $-$K. As shown in Fig.~\ref{f7}(b), $\pm$K fold to the corners of the smaller hexagonal BZ. Apparently, the nearest-neighboring corners are symmetric about the $\rm{\sigma_v}$. Thus, the valley states originating from the MoTe$_2$ monolayer stay degenerate in the presence of type-\uppercase\expandafter{\romannumeral2} AFM proximity effect.\par

In the type \uppercase\expandafter{\romannumeral3}, the valley splitting is predicted to be induced in the absence of 2 aforementioned special symmetries which ensure the valley degeneracy. Because there does not exist any symmetry that reverts the spin and momentum simultaneously, the two valley states keep irrelevant to each other no matter which points $\pm$K fold to. The numerical results verifies the symmetry analyses above. A small valley splitting $\backsim1$ meV in the valence band is extracted from the TB model and the DFT calculation as shown in Fig.~\ref{f5}(c) and (f). The main results are summarized in Table.~\ref{result}.\par
\begin{figure}
\includegraphics[width=17cm]{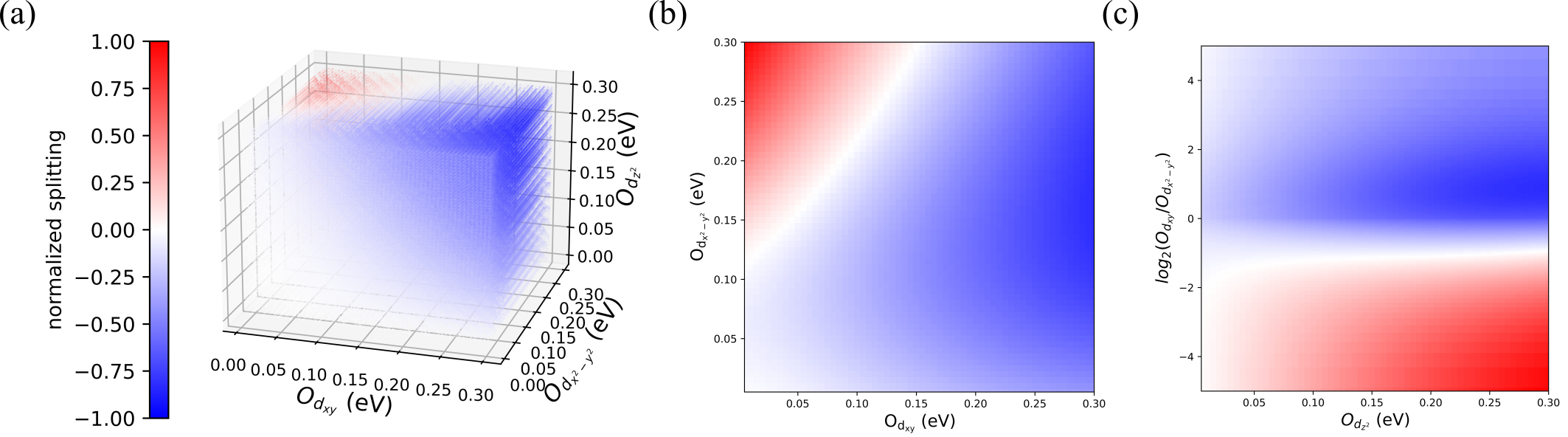}%
\caption{The valley splitting with orbital-dependent energy shift from the TB model. The splitting of the two valleys ($\Delta_{tvb}=E_{tvb}(\mathrm{K},\downarrow)-E_{tvb}(-\mathrm{K},,\uparrow)$) is normalized by its maximum 5.7 meV at $O = (0.300, 0.005, 0.300)$ eV. (b) The valley splitting with $O_{d_{xy}}$ and $O_{d_{x^2-y^2}}$. $O_{d_{z^2}}$ is fixed at 0.300 eV. (c) The valley splitting with $O_{d_{z^2}}$ and log$_2(O_{d_{xy}}/O_{d_{x^2-y^2}})$.}
\label{TBresults}
\end{figure}

\begin{table}[ht]
 \caption{Main results of the type \uppercase\expandafter{\romannumeral1}--\uppercase\expandafter{\romannumeral3} TMD/AFM hybrids.}
 \label{result}
\centering
\begin{tabular}{llll}
\hline \hline
 &Special & Kramers' & Valley\\
 &symmetry & degeneracy  & degeneracy\\
\hline Type \uppercase\expandafter{\romannumeral1}  & $\rm{\mathcal{T}t_R}$ &\ preserved &\ preserved\\
Type \uppercase\expandafter{\romannumeral2}  &\ \ $\rm{\sigma_v}$ &\ preserved &\ preserved\\
Type \uppercase\expandafter{\romannumeral3}  &\ none &\ broken &\ broken\\
\hline \hline
\end{tabular}
\end{table}
The dependence of the valley splitting in the AFM type \uppercase\expandafter{\romannumeral3} on the orbital-resolved proximity effect is further studied by tuning the modifier term in the TB model. The 3 matrix elements of $\vec{O}$ change from 0.005 to 0.300 eV independently. The valley splitting in Fig.~\ref{TBresults} reaches 5.7 meV at $\vec{O} = (0.300, 0.005, 0.300)$ eV and $-$4.7 meV at $\vec{O}=(0.300, 0.300, 0.160)$ eV. The nearly zero valley splitting appears in the region where the orbital energy shift is nonzero as shown in Fig.~\ref{TBresults}(b)--(c). It is also found that a larger energy shift of $d_{z^2}$ will lead to a more significant valley splitting when the valley splitting is nonzero.\par

\subsection{\label{3B} The mechanism of the valley degeneracy/splitting}
In order to reveal the mechanism of the valley degeneracy/splitting in the 3 AFM types, the non-degenerate perturbation is adopted. Following the perturbation treatment, the dependence of the valley splitting on the orbital-resolved energy shift will be explained with the eigenstates extracted from the TB model.\par

\begin{figure}
\includegraphics[width=17cm]{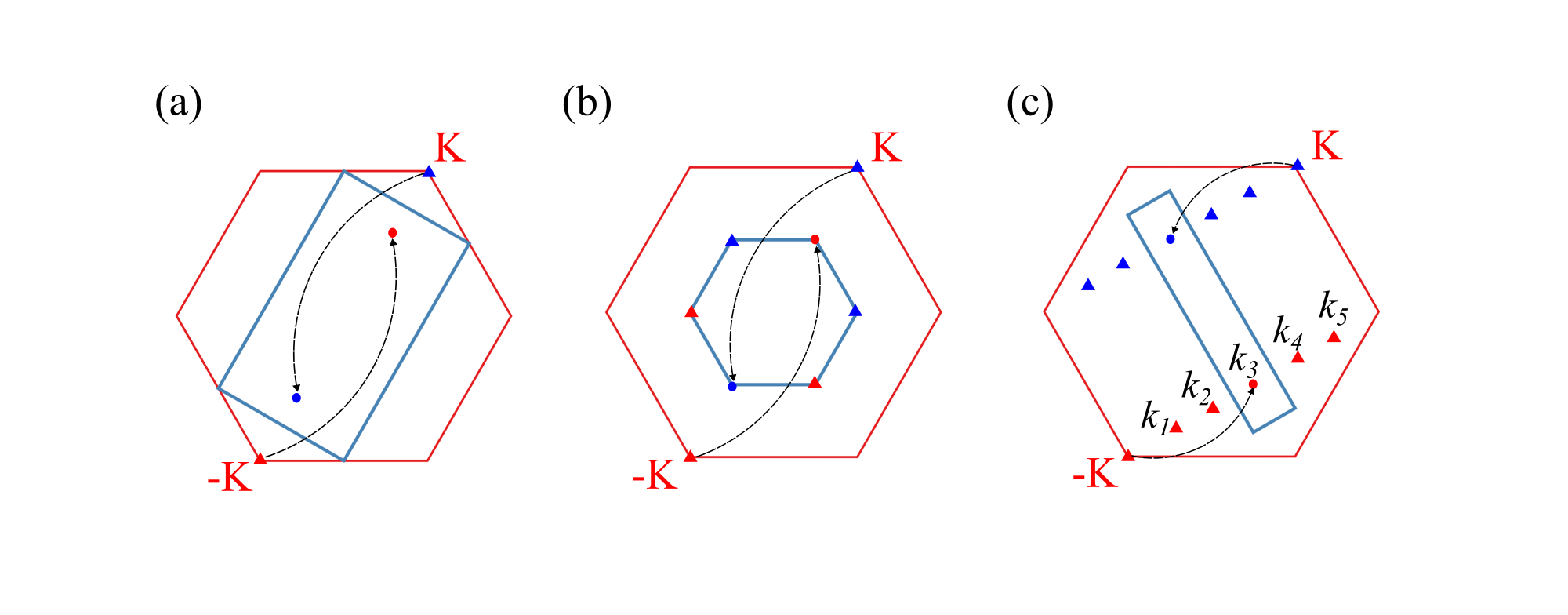}%
\caption{The PBZ of the MoTe$_2$ monolayer and the SBZ of the three SCs. The sites which the points fold to in the SBZ are denoted by colored circles. The points folding to the same sites that the K or $-$K folds to are denoted by colored triangles. (a) The SBZ of the AFM type \uppercase\expandafter{\romannumeral1}. Each $k$ point in the SBZ corresponds to 2 points in the PBZ. (b) The SBZ of the AFM type \uppercase\expandafter{\romannumeral2}. Each $k$ point in the SBZ corresponding to 4 points in the PBZ. (c) The SBZ of the AFM type \uppercase\expandafter{\romannumeral3}. $-$K (K) and the other five inequivalent $k$ points in the PBZ are folded to the $k_3$ ($-k_3$) in the SBZ.}
\label{f7}
\end{figure}

The non-degenerate perturbation starts from the construction of the atomic basis consisting of Mo $\{d_{z^2},\ d_{xy},\ d_{x^2}\}$. In the supercell (SC) composed of the multiple identical primitive cells (PCs), the eigen state is equivalent to that of the PC with the corresponding band and momentum index. The relationship between the momentum of the $n_i^{th}$ band of the PC and the $m^{th}$ band of the SC is given by
\begin{equation}
    \vec{k}_i+\vec{G}_i\rightarrow\vec{\kappa}_i;\ |k_i, n_i\rangle\rightarrow|\kappa, m\rangle.
    \label{eq3}
\end{equation}
$\vec{G}_i$ is the uniform reciprocal translation through which the $n_i^{th}$ band indexed by $k_i$ in the Brillouin zone of the primitive cell (PBZ) folds to the $m^{th}$ band indexed by $\kappa$ in the Brillouin zone of the supercell (SBZ). Considering that the PC only includes one Mo atom, the unperturbed basis of the SC is expressed as
\begin{widetext}
\begin{equation}
\begin{split}
|k_i,n_i\rangle_{\sigma}\longrightarrow|\kappa_i,m\rangle_{\sigma}=&\frac{1}{\sqrt{N_p}}\sum_{\vec{R}_p}e^{i\vec{k}_i\cdot\vec{R}_p}\sum_{\alpha}c_{\alpha,\sigma}^{n_i}(\vec{k}_i)|\alpha\rangle_{\sigma}\\
=&\frac{1}{\sqrt{N_S}}\sum_{\vec{R}_S}e^{i\vec{k}_i\cdot\vec{R}_S}\frac{1}{\sqrt{N_{\mu}}}\sum_{\alpha}\sum_{\mu}e^{i\vec{k}_i\cdot\vec{R}_{\mu}}c_{\alpha,\sigma}^{n_i}(\vec{k}_i)|\alpha,\mu\rangle_{\sigma}.
\end{split}
\end{equation}
\end{widetext}
$N_p$, $N_S$, $N_{\mu}$ is the number of the PCs, SCs and PCs per one SC, respectively. The corresponding position is denoted by $\vec{R}_p$, $\vec{R}_S$ and $\vec{R}_{\mu}$. The orbital and spin are denoted by $\alpha$ and $\sigma$. Noted that the non-degenerate perturbation to the valley states is carried out due to the decoupled spin and orbital space. The 1$^{\mathrm{st}}$ order perturbation to the energy is expressed as
\begin{equation}
\begin{split}
E^{(1)}_{n_i,\sigma}(\vec{k}_i=\vec{K})&=\langle\kappa,m|\hat{O}|\kappa,m\rangle_{\sigma}\\
&=\frac{1}{N_{\mu}}\sum_{\alpha}\sum_{\mu}\{|c^{n_i}_{\alpha,\sigma}(\vec{K})|^2O_{\alpha,\mu,\sigma}\}.   
\end{split}
\label{1stperturb}
\end{equation}
The Eq.~\ref{1stperturb} yields the zero correction because the anti-ferromagnetism leads to the zero total energy shift over the whole SC. On the contrary, the nonzero 1$^{\mathrm{st}}$ order correction is obtained through the FM or ferri-magnetic proximity effect, where the net magnetic moment is nonzero. In the case of the 2$^{\mathrm{nd}}$ order perturbation, the energy correction including the eigenstates indexed by different momentum in the PBZ may be nonzero due to the downsized BZ and the proximity effect. It is given by
\begin{equation}
\begin{split}
   &\sum_{n'}E_{n_in',\sigma}^{(2)}(\vec{k_i}=\vec{K},\vec{k}_{i'};\vec{\kappa}_i)\\
   =&\sum_{n'}\frac{1}{N_{\mu}^2}\sum_{\alpha}\frac{|c_{\alpha,\sigma}^{n'}(\vec{k}_{i'})^{*}c_{\alpha,\sigma}^{n_i}(\vec{k}_i)\sum_{\mu}e^{i(\vec{K}-\vec{k}_{i'})\cdot\vec{R}_{\mu}}O_{\alpha, \mu, \sigma}|^2}{E_n-E_{n'}} 
\end{split}
\label{2ndperturb}
\end{equation}
with $E_n\neq E_{n'}$ and $\vec{k}_{i'}+\vec{G}_{i'}=\vec{\kappa}_i=\vec{K}+\vec{G}_{i}$. For the 3 AFM types in Fig.~\ref{f2}, all the $k$ points involved in the perturbation are shown in Fig.~\ref{f7}. In the AFM type \uppercase\expandafter{\romannumeral3}, $-$K (K) corresponds to the other 5 points folding to the $k_3$ ($-k_3$) as shown in Fig.~\ref{f7}(c). Noted that for the correction $\sum_{n''}E_{n_jn'',\bar{\sigma}}^{(2)}(\vec{k}_j=-\vec{K},\vec{k}_{j''};\vec{\kappa}_j=-\vec{\kappa})$, the sum over $n''$ can be done with $\vec{k}_{j''}=-\vec{k}_{i'}$. Since $|-\vec{k}_{i'},n'\rangle_{\bar{\sigma}}$ is related to $|\vec{k}_{i'},n'\rangle_{\sigma}$ with the $\mathcal{T}S$, the sign and the absolute value of the 2$^{\mathrm{nd}}$ order energy correction are the same for the two valley states. Thus, the valley degeneracy still holds. Up to the 3$^{\mathrm{rd}}$ order, the expression is given by
\begin{widetext}
\begin{equation}
\begin{split}
  E^{(3)}_{n_i,\sigma}(\vec{k}_i=\vec{K})&=\sum_{p\neq n_i}\sum_{l\neq n_i}E_{n_il}E_{lp}E_{pn_i}\\
  &=\sum_{p\neq n_i}\sum_{l\neq n_i}\frac{1}{E_{n_i}-E_l}\frac{1}{E_{n_i}-E_p}\langle K|\hat{O}|l\rangle_{\sigma}\langle l|\hat{O}|p\rangle_{\sigma}\langle p|\hat{O}|K\rangle_{\sigma}.  
\end{split}
\label{3rdperturb}
\end{equation}
\end{widetext}
One of the brakets in Eq.~\ref{3rdperturb} is expressed as
\begin{equation}
\langle l|\hat{O}|p\rangle_{\sigma}=\frac{1}{N_{\mu}}\sum_{\alpha}c_{\alpha,\sigma}^{l}(\vec{k}_{l})^{*}c_{\alpha,\sigma}^p(\vec{k}_p)\sum_{\mu}e^{i(\vec{k}_p-\vec{k}_l)\cdot\vec{R}_{\mu}}O_{\alpha,\mu,\sigma}.
\end{equation}
In the case of the AFM type \uppercase\expandafter{\romannumeral1} or type \uppercase\expandafter{\romannumeral2}, the symmetry transformation $S$ ($S=\mathcal{T}t_{\vec{R}}$ or $\rm{\sigma_v}$) ensures the following relationship between the states with opposite momentum and spin,
\begin{equation}
\begin{split}
\langle l|\hat{O}|p\rangle_{\sigma}=&\langle l|S^{\dagger}S\hat{O}S^{\dagger}S|p\rangle_{\sigma}\\
=&\langle l_S|\hat{O}|p_S \rangle_{\bar{\sigma}},
\end{split}
\label{3rdTransStates}
\end{equation}
where $|l_S \rangle_{\bar{\sigma}}$ and $|p_S \rangle_{\bar{\sigma}}$ are transformed states. It can be inferred from Eq.~\ref{3rdTransStates} that each pair of states (e.g., $|l \rangle_{\sigma}$ and $|l_S \rangle_{\bar{\sigma}}$) participating in the unfolding process in Fig.~\ref{f7} contributes same energy correction to the corresponding modified valley state in the AFM type \uppercase\expandafter{\romannumeral1} and type \uppercase\expandafter{\romannumeral2}. Specifically, states transform as $\hat{\mathcal{T}}|l \rangle_{\sigma}=|-l \rangle_{\bar{\sigma}}$ in the AFM type \uppercase\expandafter{\romannumeral1}. Meanwhile, the $\mathcal{T}S$ of the unperturbed system enforces $\langle l|\hat{O}|p\rangle_{\sigma}=-\langle-l|\hat{O}|-p\rangle_{\bar{\sigma}}^*$. As a result, the term $E_{n_il}E_{lp}E_{pn}+E_{ln_i}E_{pl}E_{n_ip}$ in the summation over $p$ and $l$ is zero in the AFM type \uppercase\expandafter{\romannumeral1}. Without any special symmetry, a nonzero energy shift appears at two valleys in the AFM type \uppercase\expandafter{\romannumeral3}. Unlike the 2$^{\mathrm{nd}}$ order term, the sign of the energy correction is opposite for the two valley states.\par
\begin{figure}
    \centering
    \includegraphics[width=17cm]{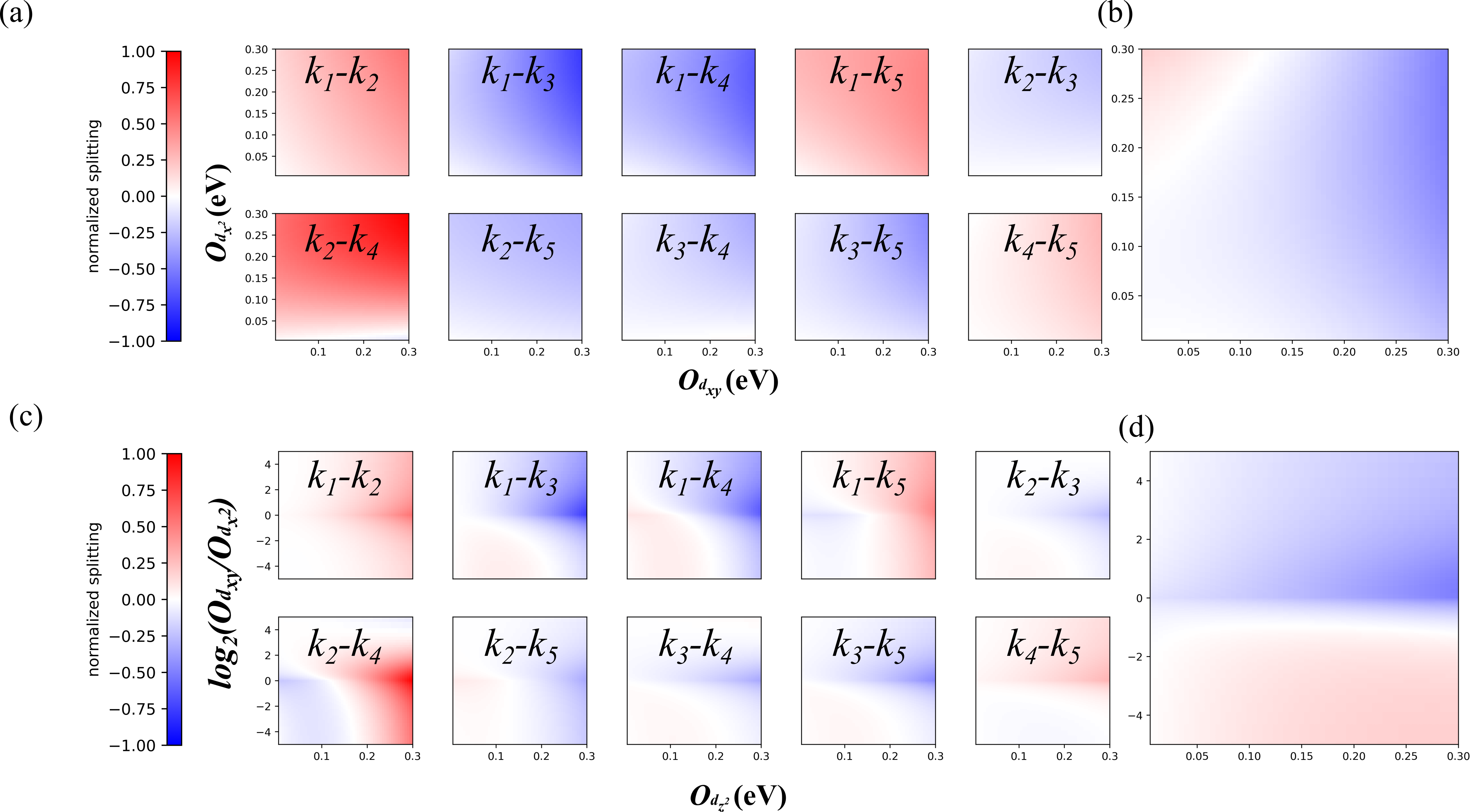}
    \caption{The simulation of the $\Delta_{tvb}$ based on the 3$^{\mathrm{rd}}$ order perturbation. The eigenstates and the eigenenergies of the folding states are extracted from the TB results, as listed in Table.~\ref{eigenstate}. The two $k$ points in (a) and (c) represent the two intermediate states. The corresponding momentum is labeled in Fig.~\ref{f7}(c). In (a) and (b), the results are normalized by the maximal absolute value of the valley splittings from the 10 paths in (a). Similarly, the normalization in (c) and (d) is carried out with the valley splittings from the 10 paths in (c). (a) The partial contribution of the 10 paths to the valley splitting with $O_{d_{xy}}$ and $O_{d_{x^2-y^2}}$. The $O_{d_{z^2}}$ is fixed to be 0.300 eV. (c) The partial contribution of the 10 paths to the valley splitting with log$_2(O_{d_{xy}}/O_{d_{x^2-y^2}})$ and $O_{d_{z^2}}$. (b), (d) The summation of the partial valley splitting in (a), (c).}
    \label{perturbPath}
\end{figure}

Based on the perturbation treatment above, the TB results in Fig.~\ref{TBresults} will be discussed as following. It is known that the eigenstates at the band edges of the valence bands are composed of the $\{d_{xy},\ d_{x^2-y^2}\}$ orbitals. Thus, it is unusual that the $\Delta_{tvb}$ of the type \uppercase\expandafter{\romannumeral3} depends on the energy shift of $d_{z^2}$ as shown in Fig.~\ref{TBresults}. In Table.~\ref{eigenstate}, the orbital components of the spin-up valence band at K, $-$K and $k_i$ ($i=1, …, 5$) are extracted through the three-band TB model. It is revealed that the large $d_{z^2}$ component is included in the other five states participating in the 3$^{\mathrm{rd}}$ order perturbation. As a result, the nominator relevant to the $d_{z^2}$ component is nonzero in Eq.~\ref{3rdperturb}. On the other hand, the failure of the 2$^{\mathrm{nd}}$ order perturbation is proved again due to the zero inner product of the $d_{z^2}$ part in Eq.~\ref{2ndperturb}. The discussion above is further validated by simulation of the energy correction contributed from 10 perturbation paths shown in Fig.~\ref{perturbPath}. The major features of the accumulative effect in Fig.~\ref{perturbPath}(b) and (d) agree well with the TB results. To sum up, the 3$^{\mathrm{rd}}$ order perturbation reveals how the valley splitting happens and depends on the orbital-resolved proximity effect.\par
\begin{figure}
    \centering
    \includegraphics[width=12cm]{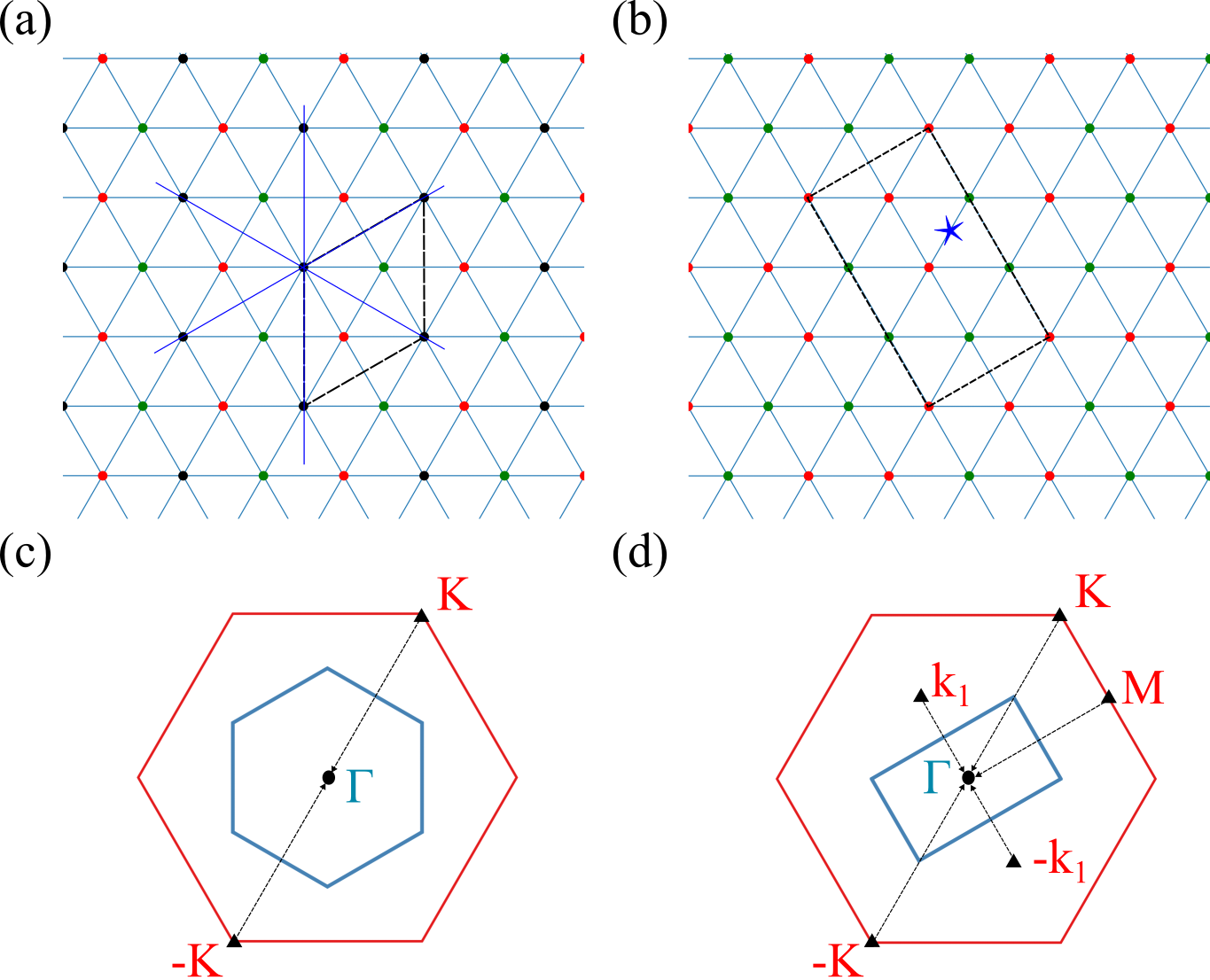}
    \caption{The structures and the corresponding BZs of the AFM type-\uppercase\expandafter{\romannumeral4} and type-\uppercase\expandafter{\romannumeral5} configurations. (a) The AFM type \uppercase\expandafter{\romannumeral4} is symmetric about the three vertical mirror planes. (b) The AFM type \uppercase\expandafter{\romannumeral5} is unchanged under the inversion combined with time-reversal symmetry. The inversion center is labeled by the blue star. (c)--(d) The BZ of AFM type-\uppercase\expandafter{\romannumeral4} and type-\uppercase\expandafter{\romannumeral5} configuration. The points folding to the $\Gamma$ are denoted by black triangles.}
    \label{failstruct}
\end{figure}
\begin{figure}
    \centering
    \includegraphics[width=17cm]{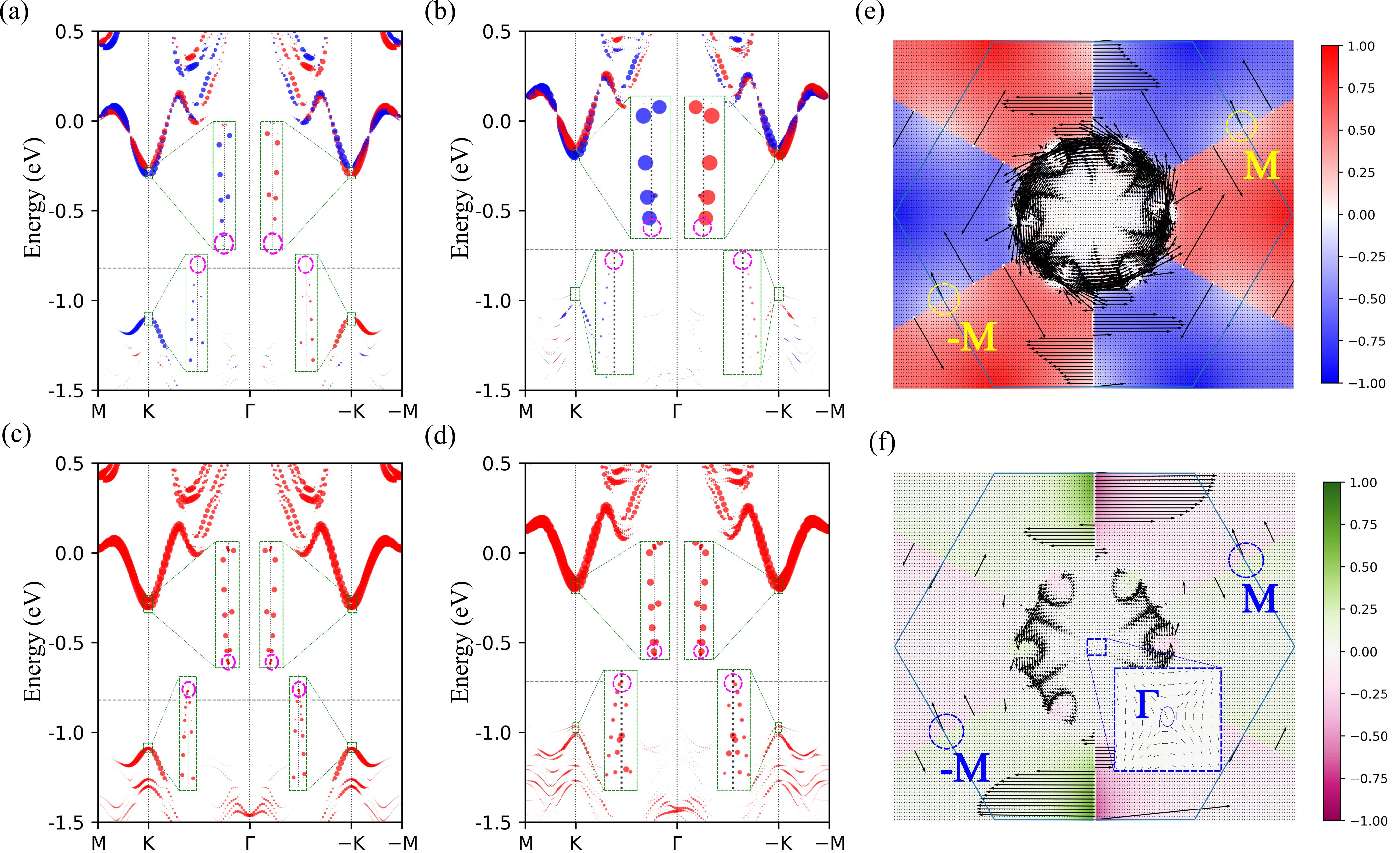}
    \caption{(a)--(d) The unfolded bands of the AFM type-\uppercase\expandafter{\romannumeral4} and type-\uppercase\expandafter{\romannumeral5} configurations. (a)--(b) The size and the shade of the red (blue) circles denote the unfolded weight factorized by the weight of $\{d_{z^2},\ d_{xy},\ d_{x^2-y^2}\}$ with positive (negative) spin $z$ component. (c)--(d) The size and the shade of the red circles denote the unfolded weight multiplied by the projection on $\{d_{z^2},\ d_{xy},\ d_{x^2-y^2}\}$. (e)--(f) The spin-textures of the projected top valence band of the MoTe$_2$ monolayer in the reciprocal space. The shade of the colors denotes the projection weight of the specific orbitals with spin $z$ component. The length of the arrow denotes the size of the projection weight of the specific orbitals with in-plane spin components. (e) The orbitals for projection are Mo $\{d_{xy},\ d_{x^2-y^2}\}$. M and $-$M which participate in the degenerate perturbation in the type \uppercase\expandafter{\romannumeral5} are labeled by yellow circles. (f) The orbitals for projection are Te $p_{x}$. The region centered at $\Gamma$ is magnified with the blue dashed frame, where the nonzero intra-plane spin component at $\Gamma$ is labeled by blue circle. The spin-texture projecting on Te $p_{y}$ reflects similar property, only different by the orientation of the spin. Thus, the projection on $p_{y}$ is not included.}
    \label{failBand}
\end{figure}

\begin{table}[ht]
\centering
\caption{The spin-up eigenstates of the valence band within the three-band basis at 7 points of the PBZ. The eigenstate of the valence band at $-$K is also listed. The zero energy is fixed at the maximum of the valence band.}
\begin{tabular}{l|lll|l}
\hline \hline
 &$|d_{z^2},\uparrow\rangle$ & $\ \ \ |d_{xy},\uparrow\rangle$  & $|d_{x^2-y^2},\uparrow\rangle$ &energy (eV)\\
\hline $|\mathrm{K},\uparrow\rangle$  & 0.000 &-0.707 &\ \ \ \ \ \ \ \ \ \ 0.707i& -0.214\\
$|k_1,\uparrow\rangle$  &0.654 &\ 0.278+0.368i &-0.541+0.261i & -0.404\\
$|k_2,\uparrow\rangle$  &0.856 &\ 0.013+0.023i &-0.446+0.261i & -0.361\\
$|k_3,\uparrow\rangle$  &0.880 &-0.294\ -0.166i &-0.170+0.287i & -0.312\\
$|k_4,\uparrow\rangle$  &0.856&-0.379\ -0.237i &\ 0.234+0.111i & -0.361\\
$|k_5,\uparrow\rangle$  &0.654 &-0.330\ -0.410i &\ 0.511\ -0.188i & -0.404\\
$|-\mathrm{K},\uparrow\rangle$  & 0.000 &\ 0.707 &\ \ \ \ \ \ \ \ \ \ 0.707i& \ 0.000\\
\hline \hline
\end{tabular}
\label{eigenstate}
\end{table}
It should be noted that the well-defined valley states will no longer exist if the two valley points K and $-$K fold to the same point $\Gamma$. The AFM type-\uppercase\expandafter{\romannumeral4} and type-\uppercase\expandafter{\romannumeral5} configurations are constructed as the negative examples. In both types, the two valley states fold to $\Gamma$ as shown in Fig.~\ref{failstruct}. According to the TB model, the valley degeneracy and valley splitting are predicted in the AFM type \uppercase\expandafter{\romannumeral4} and type \uppercase\expandafter{\romannumeral5} respectively. However, the nearly zero weight of band projection on spin $z$ component of $\{d_{z^2},\ d_{xy},\ d_{x^2-y^2}\}$ occurs at K and $-$K in the DFT calculation as shown in Fig.~\ref{failBand}(a)--(b). Meanwhile, the weight of band projection on $\{d_{z^2},\ d_{xy},\ d_{x^2-y^2}\}$ at two valleys is nonzero in Fig.~\ref{failBand}(c)--(d). It indicates that the two valley states $|\mathrm{K},\downarrow\rangle$ and $|-\mathrm{K},\uparrow\rangle$ mixes with each other after the introduction of the AFM proximity effects in Fig.~\ref{failstruct}(a)--(b). In order to explain the anomalous results, the degenerate perturbation is adopted. The interested subspace is spanned by two valley states $|\mathrm{K},\downarrow\rangle$ and $|-\mathrm{K},\uparrow\rangle$. Up to the 2$^{\mathrm{nd}}$ order degenerate perturbation through Lowdin partitioning equation \cite{lowdingmethod}, the off-diagonal matrix element between two valley states is expressed as
\begin{equation}
 \lambda = H_{K,-K}^{(2)}=\frac{1}{2}\sum_l\hat{O}_{K,k_l}\hat{O}_{k_l,-K}[\frac{1}{E_K-E_l}+\frac{1}{E_{-K}-E_l}].
\label{degpertub}
\end{equation}
$\lambda$ will become nonzero under two conditions. First, there are points other than K and $-$K folding to the $\Gamma$. Second, common orbital components exist in $\{|{k}_l\rangle,-|{k}_l\rangle\}$ and $\{|\mathrm{K},\downarrow\rangle, |-\mathrm{K},\uparrow\rangle\}$. These orbital components of the $\{|{k}_l\rangle,-|{k}_l\rangle\}$ are further required to include both spin channels. Then, the eigenstates in the subspace spanned by two valley states will be in the form of $|\mathrm{K},\downarrow\rangle+\frac{\lambda^*}{|\lambda|}|-\mathrm{K},\uparrow\rangle$ no matter how small the $\lambda$ is. Consequently, the expectation value of the spin operator is zero with arbitrary spin-quantization axis.\par

Based on the analysis above, the spin-texture of the top valence band in MoTe$_2$ monolayer is extracted to check the orbital and spin components. As shown in the yellow circled region of Fig.~\ref{failBand}(e), the projection weight of the $\{d_{xy},\ d_{x^2-y^2}\}$ with in-plane spin direction is nonzero at M. Although $\{d_{xy}, d_{x^2-y^2}\}$ component becomes zero at $\Gamma$, a tiny but nonzero projection on Te $\{p_x, p_y\}$ has been found at $\Gamma$ as shown Fig.~\ref{failBand}(f). The finite in-plane spin component means both spin-up and spin-down channels are contained for the projected orbitals. Thus, the folded state $|\mathrm{M}\rangle$ drives the mixture of the two valley states in the AFM type \uppercase\expandafter{\romannumeral5}. In the type \uppercase\expandafter{\romannumeral4}, the mixed states are caused by the folded state $|\Gamma\rangle$. That’s why the nearly zero weight of projection on Mo$\{d_{z^2},\ d_{xy},\ d_{x^2-y^2}\}$ with spin $z$ component occurs in the DFT calculation of the AFM type \uppercase\expandafter{\romannumeral4} and \uppercase\expandafter{\romannumeral5}. The problem is naturally avoided in the three-band TB model because the spin space is decoupled with the orbital space, which excludes the degenerate perturbation. Thus, the construction of the SC where K and $-$K fold to the $\Gamma$ should be avoided since the 2$^{\mathrm{nd}}$ order degenerate perturbation couples the well-defined two valley states.\par
\subsection{\label{3C} The effect of the interlayer distance and $U_{\mathrm{eff}}$ on the valley splitting}
Although the hybrid systems are novel due to the unstable MnO (111) monolayer, instructive results are also provided for engineering functional materials. In order to investigate the method of controlling the valley splitting, the effects of the $d$ and $U_{\mathrm{eff}}$ on the valley splitting have been studied through the DFT calculations. In spite of influence from the magnetic configuration, only small difference less than 0.1 \r{A} is found in the relaxed FM type, AFM type \uppercase\expandafter{\romannumeral1} and type \uppercase\expandafter{\romannumeral3} with $d$ = 4.754, 4.807, 4.727 \r{A} respectively under $U_{\mathrm{eff}}$ = 4 eV. The AFM type \uppercase\expandafter{\romannumeral2} is not included because of the additional effect of atomic substitution. Thus, the magnetic configuration effect is neglected in the following calculations.\par

\begin{figure}
    \centering
    \includegraphics[width=13cm]{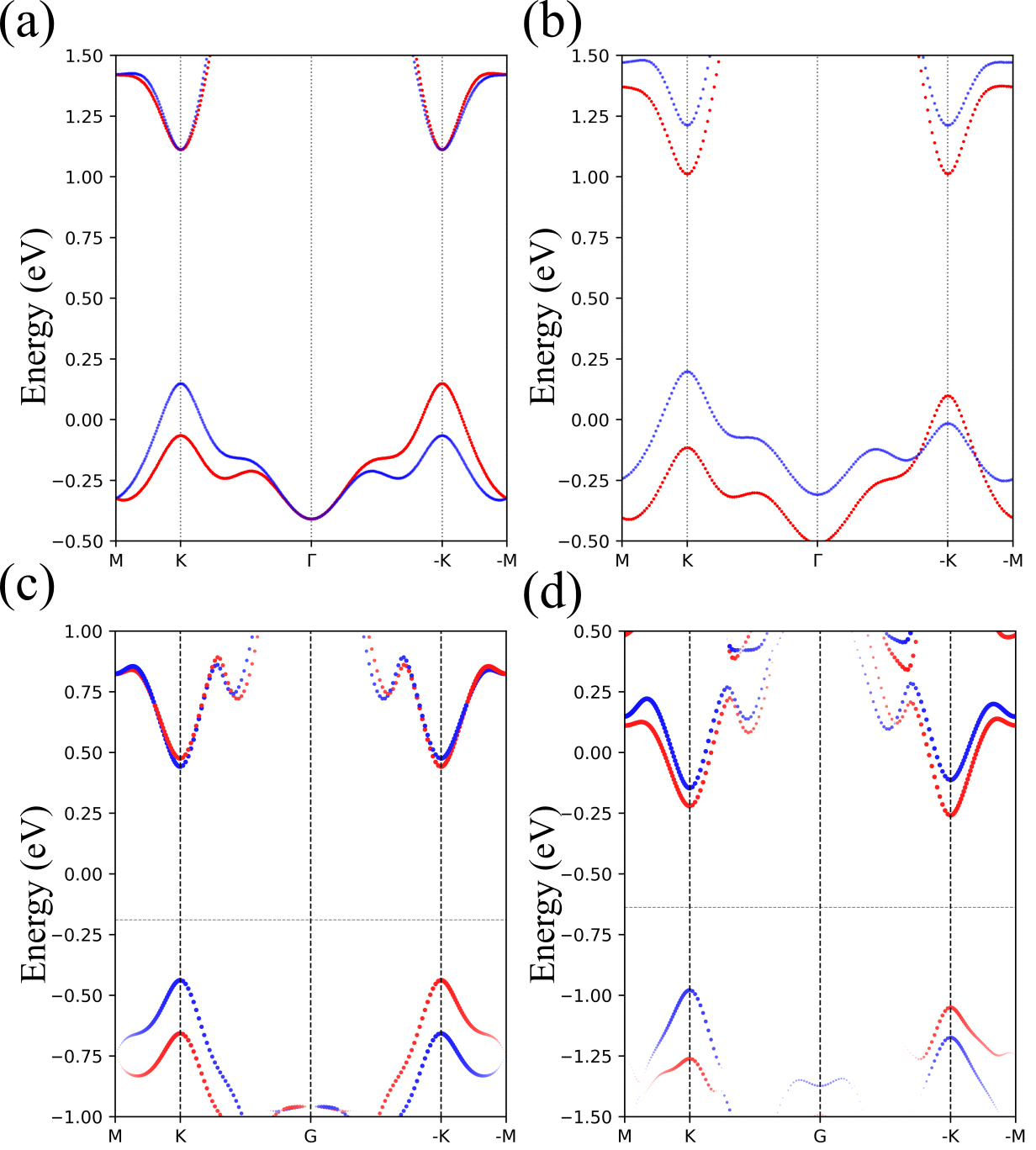}
    \caption{(a)--(b) The band structures from the TB model. The onsite energy shift is zero in (a) and $\vec{O}=-(0.100,0.050,0.050)$ eV in (b). Red/Blue denotes the spin-up/spin-down channel. (c)--(d) The band structures from the DFT calculations of the freestanding MoTe$_2$ monolayer in (c) and the MoTe$_2$/MnO hybrids in (d). The magnetic moment of the Mn$^{2+}$ aligns along the direction normal to the plane. Red/Blue circles represent the projected bands. The shade denotes the weight of projection on Mo $\{d_{z^2},d_{xy},d_{x^2-y^2}\}$ with spin-up/spin-down.}
    \label{fmtbDFTband}
\end{figure}

\begin{figure}
    \centering
    \includegraphics[width=13cm]{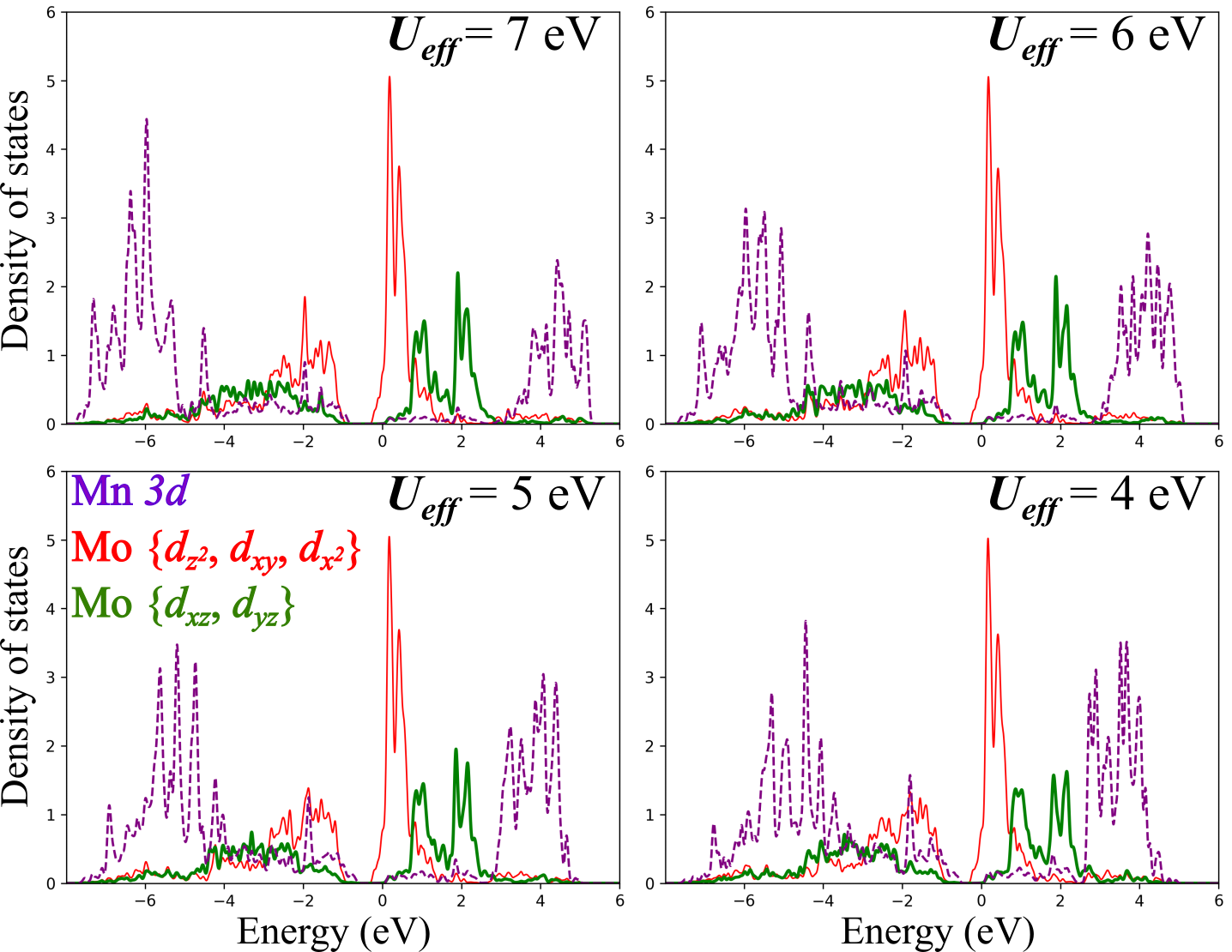}
    \caption{The PDOS on Mn 3$d$ orbitals and Mo 4$d$ orbitals with $d$ = 3.6 \r{A} under different $U_{\mathrm{eff}}$.}
    \label{PDOS}
\end{figure}

The properties of the FM types are discussed first. As listed in Table.~\ref{tab3}, the top valence band splitting (vs$_v$) and the lowest conduction band splitting (vs$_c$) at two valleys tend to decrease as the $U_{\mathrm{eff}}$ increases. The vs$_c$ keeps decreasing while the vs$_v$ decreases until the sign flips when the $d$ increases from 4.4 to 4.8 \r{A}.
\begin{table}
    \centering
        \caption{The top valence band splitting vs$_v$ and the lowest conduction band splitting vs$_c$ of the FM type under different $U_{\mathrm{eff}}$ values. The splittings are in the unit of meV. The atomic sites of the oxygen in each structure with fixed Mn-Mo distance under specific $U_{\mathrm{eff}}$ have been optimized.}
\begin{tabular}{l|ll|ll|ll|ll}
\hline \hline
$U_{\mathrm{eff}}$ (eV)& \multicolumn{2}{|l|}{7}&\multicolumn{2}{|l|}{6}& \multicolumn{2}{|l|}{5}& \multicolumn{2}{|l}{4}\cr
\hline
$d$ (\r{A})& vs$_{v}$ & vs$_{c}$ & vs$_{v}$ & vs$_{c}$  & vs$_{v}$ & vs$_{c}$ & vs$_{v}$ & vs$_{c}$\cr
\hline \hline
 3.6 & 71.3 & 112.2 & 79.3 & 122.9 & 87.8 & 135.0 & 96.6 &148.5 \cr
 4.0 & 33.4 & \ 60.5 & 38.0 & \ 65.6 & 43.5 & \ 71.4 & 50.3 &\ 79.7 \cr
 4.4 & \ 7.4 & \ 27.8 & \ 8.9 & \ 29.8 & 11.1 & \ 32.5 & 16.1 &\ 40.6 \cr
 4.8 & -4.0 & \ 11.8 & -3.9 & \ 13.0 & -2.3 & \ 15.3 & \ 2.8 &\ 21.3 \cr
\hline \hline
\end{tabular}
\label{tab3}
\end{table}
Noted that the proximity effect results from the overlap between the Mo 4$d$ orbitals and Mn 3$d$ orbitals. Consequently, the Mo orbitals with spin parallel to the magnetic moment of the Mn tend to be lowered. As shown in Fig.~\ref{fmtbDFTband}, the spin-down projected bands lower while the spin-up ones raise compared to the case of freestanding monolayer in Fig.~\ref{fmtbDFTband}(b). It is consistent with the bands from the TB model in Fig.~\ref{fmtbDFTband}(c) where the onsite energy shift is negative for spin-up channel.
As a result, a larger overlap implies a larger magnetization of Mo, which means a larger valley splitting. When $U_{\mathrm{eff}}$ decreases, Mn 3$d$ orbitals tend to be more delocalized. Then, the magnetization of the Mo 4$d$ orbitals strengthens due to the larger overlap from the more extensive 3$d$ wavefunction. It is reflected from the enhanced hybridization between the Mn 3$d$ orbitals and Mo 4$d$ orbitals through the projected density of states (PDOS) in Fig.~\ref{PDOS}. Similarly, the valley splitting of the FM type decreases as the $d$ increases from 3.6 to 4.4 \r{A} with fixed $U_{\mathrm{eff}}$. When the $d$ up to 4.8 \r{A}, the magnetization from the direct overlap is significantly suppressed. On the other hand, the superexchange-like process (SE) originating from the kinetic energy of electrons lowers the Mo 4$d$ orbitals with the spin anti-parallel to the magnetic moment of Mn. The mechanism of anti-ferromagnetism can be physically described by the virtual hopping process involving the Mo 4$d$ and the Mn 3$d$ orbitals via S orbitals \cite{superexchange}. Such high-order term from the perturbation scales as $\frac{t_{\mathrm{eff}}^2}{U}$. The effective hopping $t_{\mathrm{eff}}=\frac{t_{dp}^{\mathrm{Mn}}t_{dp}^{\mathrm{Mo}}}{\Delta}$ contains the charge transfer energy $\Delta$, the hopping term for Mn 3$d$-Te $p$ and the hopping term for Mo 4$d$-Te $p$. Thus, the magnetization from the SE is negligible in the case of small $d$ compared to that from the direct overlap. Noted that the SE decreases much more slowly than direct overlap due to the fixed Mo-S and slowly enlongated Mn-S distances as $d$ increases. When the increasing $d$ is large enough, the direct overlap decays exponentially and tends to be zero. The AFM SE subsequently dominates the magnetization of Mo. That’s why the vs$_v$ becomes negative in the case of $d$ = 4.8 \r{A} for $U_{\mathrm{eff}}=7, 6, 5$ eV. By comparison, the vs$_c$ remains positive. The difference results from the specific orbital components. As the main orbital component of the lowest conduction band at valleys, Mo $d_{z^2}$ enables more direct overlap than the in-plane distributed Mo $\{d_{xy},\ d_{x^2}\}$ orbitals.\par
Analogous to the discussion above, the vs$_v$ in the AFM type \uppercase\expandafter{\romannumeral3} shares the similar dependence on the $d$ and $U_{\mathrm{eff}}$. The vs$_v$ of the type \uppercase\expandafter{\romannumeral4} is extracted under different $U_{\mathrm{eff}}$. Only the cases with $d$ = 3.6 \r{A} are shown because the splittings in other insulating cases are smaller by nearly an order of magnitude. As listed in Table.~\ref{tab4}, vs$_v$ increases with the decreasing $U_{\mathrm{eff}}$. According to the analysis in Sec.~\ref{3B}, the size of the vs$_v$ partially depends on the absolute value of each element in $\vec{O}$. Meanwhile, these spin-up orbital-resolved energy shifts in the AFM type \uppercase\expandafter{\romannumeral3} are the same as those in the FM type. It is ensured by the aforementioned construction of the AFM configurations. Consequently, a similar trend of the $U_{\mathrm{eff}}$-dependent valley splitting in the FM type is expected to appear in the AFM type \uppercase\expandafter{\romannumeral3} as shown in Table.~\ref{tab4}. With the discussion above, tuning the magnetization of Mo is the key to manipulate the valley splitting in both FM and AFM types. Controllable magnetization can be obtained through interlayer distance and suitable magnetic layers with extensive or localized orbitals in real materials.\par

\begin{table}
    \centering
        \caption{The top valence band splitting vs$_v$ with $d$ = 3.6 \r{A}.}
\begin{tabular}{l|l|l|l|l}
\hline \hline
$U_{\mathrm{eff}}$ (eV)& 7 & 6 & 5 &4\cr
\hline
vs$_v$ (meV)& -0.32 & -0.43 & -0.64 &-0.99\cr
\hline \hline
\end{tabular}
\label{tab4}
\end{table}

\section{Conclusion}
A three-band tight-binding model is extended to investigate the proximity effect of the intra-plane AFM configurations. Two specific AFM types are predicted to preserve valley degeneracy resulting from the $\rm{\mathcal{T}t_R}$ or $\rm{\sigma_v}$ symmetry. Another type with trivial symmetries unsurprisingly breaks the valley degeneracy. Through symmetry analysis, valley degeneracy/splitting in the 3 particular systems is well explained, which agrees with the unfolded bands from the TB model and DFT calculations. The effect of the orbital-dependent energy shift is investigated with the TB model as well. The mechanism is revealed through the non-degenerate perturbation theory. Further constraints of the SC are proposed based on the degenerate perturbation. Beyond the analytical study, the magnetization of the Mo, which depends on $U_{\mathrm{eff}}$ and interlayer distance $d$, is found to be effective to tune the valley splitting in real materials.\par

In the present paper, the extended TB model captures the main features of the composite systems within the low-energy region, which may pave the way to study the corresponding quantum states at valleys. It provides a simple but accurate enough pathway to describe the low-energy physics for valleytronics, especially in nano-scale systems. Invalid cases where the well-defined valley states disappear can be avoided with the proposed constraints of the AFM configurations. We believe that these results are instructive for flexible manipulation of valley states. With combination of valleytronics and spintronics, the valley-splitting states and valley-degenerate states can be switched by tuning the magnetic structure of the proximity layer. Furthermore, it is known that the number of AFM insulators is much more than that of FM ones. Thus, the results enormously expand the prospect of proximity for valley polarization.\par

\section*{Acknowledgments}
This work was supported by the National Key R\&D Program of China (Grant No. 2022YFA1402401) and the National Natural Science Foundation of China (Grant No. 11521404). Computational resources were supported by the Center for High Performance Computing at Shanghai Jiao Tong University.

\providecommand{\noopsort}[1]{}\providecommand{\singleletter}[1]{#1}%

\end{document}